\documentclass[journal,onecolumn]{article}
\usepackage[margin=2cm]{geometry}
\usepackage{parskip}
\PassOptionsToPackage{hyphens}{url}
\usepackage{hyperref}
\hypersetup{
	colorlinks,
	linkcolor={blue},
	citecolor={blue},
	urlcolor={blue}
}
\usepackage[utf8]{inputenc}
\usepackage[T1]{fontenc}
\usepackage{subcaption}
\usepackage{float}
\usepackage[round, authoryear, sort]{natbib}
\usepackage{lipsum}
\usepackage{graphicx}
\usepackage{caption}
\usepackage{svg}
\usepackage[affil-it]{authblk}

\title{Image background assessment as a novel technique for insect microhabitat identification}
\author[1,*]{Sesa Singha Roy}
\author[2]{Reid Tingley}
\author[1,*]{Alan Dorin}
\affil[1]{Dept. of Data Science and AI, Faculty of IT, Monash University, Clayton 3800, VIC Australia}
\affil[2]{School of Biological Sciences, Monash University, Clayton 3800, VIC, Australia}
\affil[2]{EnviroDNA, 293 Royal Parade, Parkville, VIC 3052, Australia}

\affil[*]{Corresponding authors : sesa.singharoy@monash.edu, alan.dorin@monash.edu}

\date{}

\begin{document}

\maketitle

\begin{abstract}
The effects of climate change, habitat fragmentation under increased urbanisation, industrial agriculture and land clearing, are changing the way insects occupy habitat. Some species are highly adaptable and may utilise anthropogenic microhabitat features for aspects of their existence, either because they prefer them to natural features, or because they have no choice. Other species are tenuously dependent on natural microhabitats, having to locate these within increasingly hostile landscapes. Consequently, humans are encountering insects in new settings. Identifying and analysing these insects' use of natural and anthropogenic microhabitats is important to assess their responses to a changing environment, for improving pollination and managing invasive pests. But such studies are costly and time-consuming.\\

Traditional studies of insect microhabitat use can now be supplemented by machine learning-based insect image analysis. Typically, research has focused on automatic insect classification, but valuable data appearing in image backgrounds has been ignored. In this research, we analysed the backgrounds of insect images available on the Atlas of Living Australia database to determine the microhabitats in which they are commonly photographed. We analysed the microhabitats of three globally distributed insect species that are common across Australia: Drone flies (\textit{Eristalis tenax}), European honey bees (\textit{Apis mellifera}), and European wasps (\textit{Vespula germanica}). Image backgrounds were classified as either natural or anthropogenic microhabitat using computer vision and machine learning tools benchmarked against a manual classification algorithm.\\

We found Drone flies and European honey bees to be most commonly photographed in natural microhabitats, confirming their need for natural havens within our cities. European wasps were less likely to be seen in these areas and were commonly seen in anthropogenic microhabitats. This data supports the view that these insects are well adapted to survive in the built environment, and that the management of this invasive pest requires thoughtful reduction of their access to human-provided resources. 
The assessment of insect image backgrounds is instructive to document the use of microhabitats by insects, especially those encountered in urban environments where they are commonly photographed. The method offers insight that is increasingly vital for biodiversity management as urbanisation continues to encroach on natural ecosystems and we must consciously provide resources within built environments to maintain insect biodiversity and manage invasive pests.
\newline


\end{abstract}

\setcounter{section}{0}
\setcounter{subsection}{0}

\section{Introduction}

Insects are major contributors to terrestrial biodiversity and underpin diverse ecological and economic activities. For example, they act as pollinators \citep{losey2006economic}, pests \citep{aukema2011economic}, disease vectors \citep{lounibos2002invasions}, and they are a direct food source for many organisms \citep{hallmann2017more}. Unfortunately, insects are highly vulnerable to the current rapid increase in urbanisation, something that is having major impacts on global biodiversity \citep{bidau2018doomsday, sanchez2019worldwide, cardoso2020scientists}. Urbanisation has lead to modification of insect habitats as humans make space for industry, our residences and our agriculture \citep{sanchez2019worldwide}. These changes have lead a to loss of habitat area, habitat fragmentation and increases in habitat isolation for many species \citep{zschokke2000short, zanette2005effects}. Research has highlighted a decrease in arthropod abundance by $78\%$, a reduction in species richness by $34\%$ in German grasslands between $2008-2017$ \citep{seibold2019arthropod}, a drop of more than $75\%$ in flying insect biomass over a period of $27$ years in German nature reserves \citep{hall2017city}, a $33\%$ decrease in the number of bees and hover-fly species over a span of $33$ years in Britain \citep{powney2019widespread}, and a $71\%$ decrease in native butterflies in Britain over $20$ years \citep{thomas2004comparative}. This has been labelled by some an "Insect Armageddon" \citep{leather2017ecological}. And although there are those who question the extent of the issue \citep{saunders2020moving}, there is little doubt that a potential decline in insect numbers and biodiversity, even if localised, is of serious concern and worthy of attention. 
\newline

\begin{figure}[ht]
     \centering
      \includegraphics[{width=15cm}]{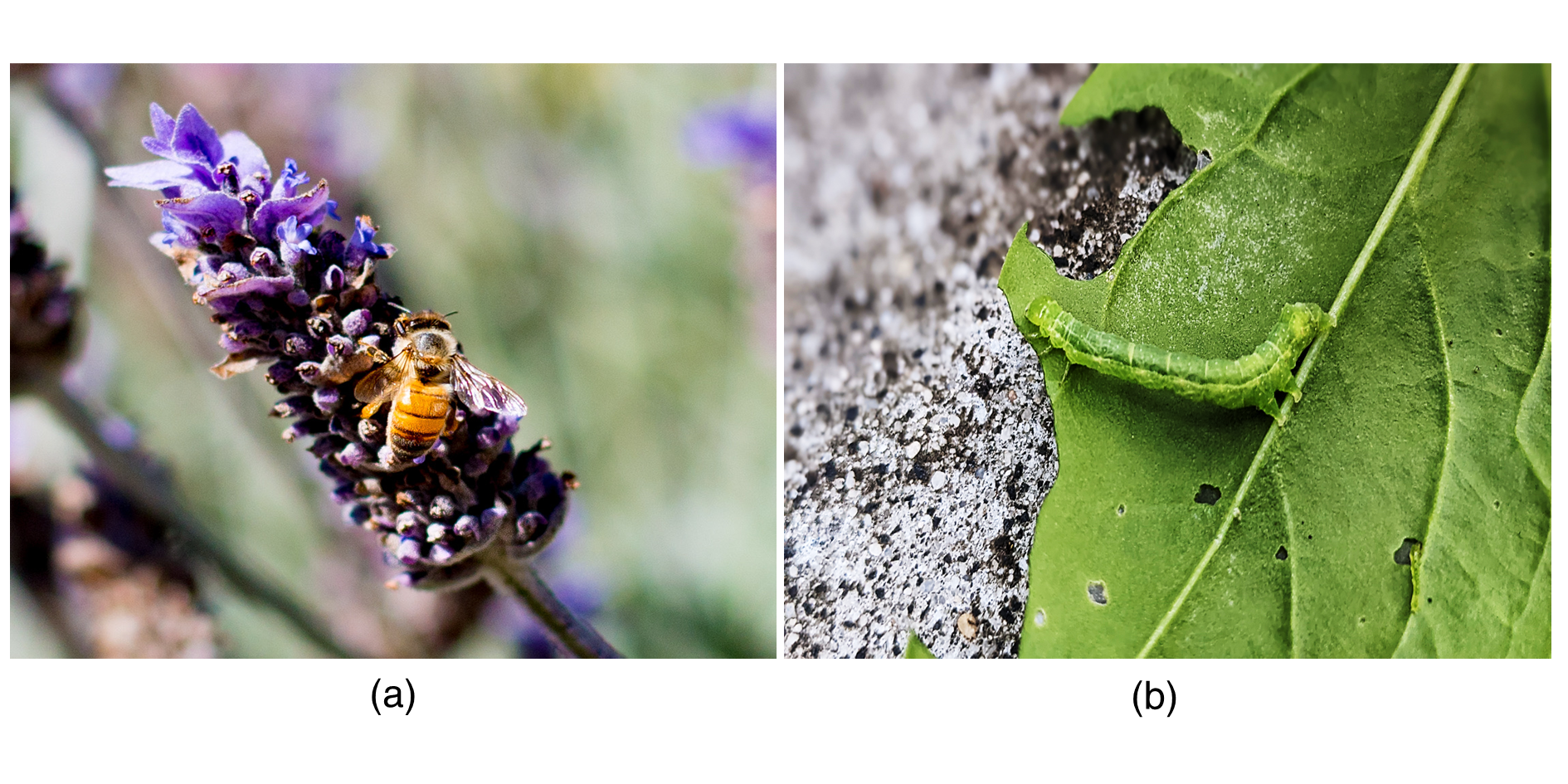}
        \caption{Insects in urban green spaces. (a) Honey bees are common garden pollinators. (b) Caterpillars can be backyard pests that consume spinach leaves and other vegetables (but their butterflies can, by contrast, be valuable pollinators). Images © Copyright SSR 2022.}
        \label{fig:insecthabitateg}
\end{figure}

Some insect species are comfortable in, or adapt to, urban habitats (Fig. \ref{fig:insecthabitateg}). In some cases, insect diversity might be high in urban refuges \citep{baranova2015waste, heneberg2016succession}. Quarries and waste-dump sites for example, provide homes for flies and soil dwelling invertebrates like beetles, spiders and harvestmen \citep{heneberg2014dry, smedt2017succession, heneberg2016succession}. Urban areas may also provide refuge to rare and threatened species such as native bees, butterflies, ground beetles and weevils \citep{jones2013invertebrates}. An increase in urban green spaces, parks, gardens, green infrastructure, even vacant wasteland, can improve the quality of life for local (human) residents and help control air pollution and the urban heat island effect \citep{ballinas2016urban, carrus2015go}. But also, the moderation of climate these areas produce, and the relative abundance of food and water, may make cities "pseudo-tropical bubbles" \citep{shochat2006patterns} that increase the abundance of some insects including ants, bees, caterpillars / butterflies and hoverflies \citep{uno2010diversity, dylewski2019all}. Such insects have therefore entered human residential areas where they are often encountered \citep{azmy2016responses}.
\newline

The microhabitats occupied by urban insects can vary based on their community structures and environmental interactions, which in turn effects local abundances and distributions. Consequently, the micro-climate in some urban spaces fosters increased abundance of some insect species but can be simultaneously detrimental to others \citep{mckinney1999biotic}. Buildings may alter locallight conditions and effect temperature and moisture in their vicinity \citep{arnfield2003two}. Urban structures in general can fragment green spaces and hinder species dispersal, resulting in a shift in species composition \citep{kozlov1996patterns, van1999habitat}. For all of these reasons, understanding the microhabitats occupied by insects can therefore be beneficial to assess insect behaviour, understand their physiology, phenology, and, importantly, to assist us to manage insects' potential responses to environmental change \citep{pincebourde2020there}. Hence, in this article we investigate the microhabitats of insects using novel image background analysis techniques applied to insect images posted online.

Insect microhabitat studies are traditionally carried out by ecologists who collect data manually from a selected study site. For example, \citep{kolenda2020deadly} executed a study in south-western Poland to identify the microhabitats of ants in wasteland. Researchers on pollinators like wild bees, butterflies and hover-flies, may use hand nets or pan traps to capture insects \citep{dylewski2019all} in different microhabitats. This is costly, time-consuming, and the expertise required presents practical obstacles  \citep{buchs2003biodiversity}. The automation of insect survey methods can potentially mitigate a shortage of human expertise. Image-based techniques for monitoring vertebrates, and invertebrate insects, have gained popularity recently \citep{yousif2019animal, norouzzadeh2018automatically, steenweg2017scaling, steen2017diel}. A variety of image types are used in these surveys including remote-sensing images \citep{hall2016remote, torresan2017forestry, zhang2019monitoring}, a combination of camera trap and remote sensing images \citep{ayres2018forest, choi2019monitoring} and even web-based images \citep{elqadi2017mapping}. Image-based techniques may require manual validation by experts \citep{preti2021insect}, but improvements in machine learning and computer vision mean that many insect studies can be automated to a large extent \citep{elqadi2017mapping, amarathunga2021methods, waldchen2018machine, joly2019overview}. Research methods that use such image-based techniques, typically, as might be expected, segment insect pixels from the remainder of the image to identify the species \citep{maharlooei2017detection, ebrahimi2017vision, qing2012insect, liu2016detection, deng2018research}. Several deep-learning based methods have been used in insect classification and monitoring of this type \citep{li2020crop, cheng2017pest, buschbacher2020image, tetila2020detection, ren2019feature}. However, as we demonstrate in this article, the image background against which the insect subject appears contains potentially valuable information. To the best of our knowledge, no research has yet analysed the backgrounds of insect images. Here we conduct such an analysis investigating clues about insect microhabitat to reveal how common species engage with their environments. Specifically, we wish to study how insects seek out natural features (like flowers, leaves, grass, sticks, soil, tree bark or fruits) or anthropogenic microhabitat features (like rubbish bins (garbage cans), food scraps, brickwork, paving stones, fence palings, fly-screens, food containers, window panes, walls, metal rods, spoons, but also human hands, human hair, clothing etc.). Our interests in these two broad classes of microhabitat utilisation is of current importance as climate change alters insect habitat availability \citep{halsch2021}, and increasing urbanisation either forces insects to leave degraded natural habitat, or offers them an opportunity to adapt to resource rich urban environments. Studies of insect microhabitat utilisation therefore improve our knowledge of insect-human encounters. They potentially enable us to shape our environment to improve insect biodiversity and reduce pest abundance.
\newline

The study aim is to investigate insect microhabitat use by both manual and automatic analysis of insect image backgrounds. We report on experiments using an analysis pipeline we developed to determine the relative extent to which three species are observed occupying natural or anthropogenic microhabitats in Australia.\\

Our study was conducted using images posted online in the Atlas of Living Australia (\url{www.ala.org}), a web-based database of Australian biodiversity. We devised a human/manual algorithm to classify insect image backgrounds as either natural or anthropogenic (Section \ref{method:manual}). Then, we devised computer vision and machine learning algorithms (Section \ref{feature_ext}, \ref{training}) to perform this same classification automatically and benchmark it against our manual method. We outline our methodology in Section \ref{methodology} and present our results in Section \ref{results} along with a discussion (Section \ref{discussion}) of strengths and weaknesses of this new research approach.

\section{Methodology} 
\label{methodology}

\begin{figure}[H] 
     \centering
     \includegraphics[{width=14cm}]{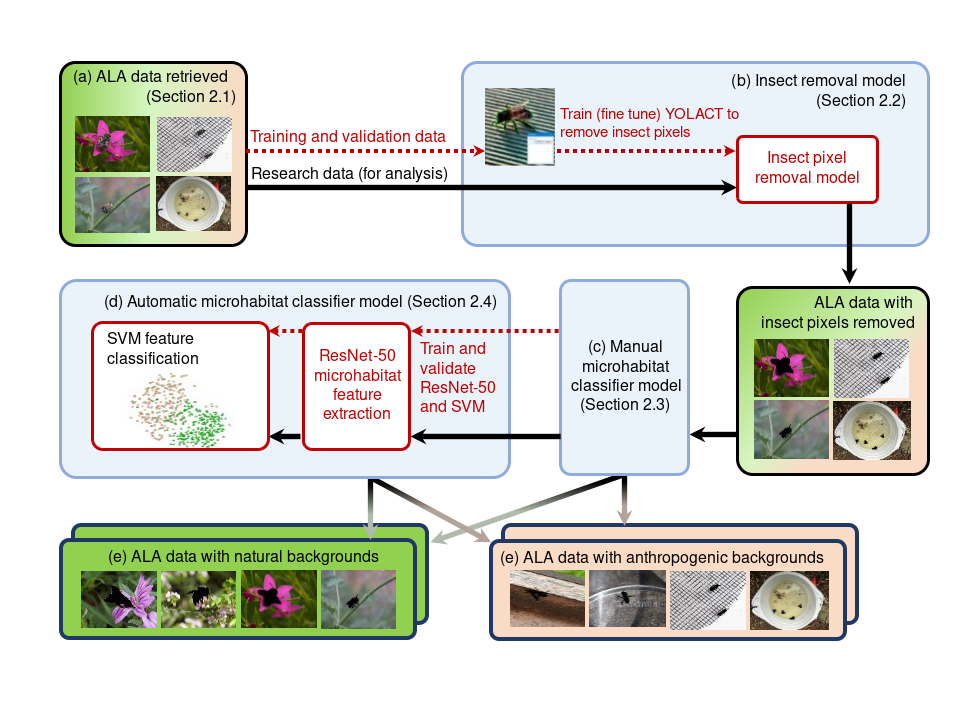}
        \caption{Overview of the methodology: (a) Insect images are collected from ALA; (b) Insect pixels are removed from the images to generate a dataset of images containing only insect backgrounds; (c) A Manual classification algorithm is used to classify the insect image backgrounds; (d) Some manually classified images are taken to create Training and validation datasets for the new automatic classifier model; Then, the automatic classifier model is trained for background classification, tested and validated; (e) The trained automatic model isused to classify the background-only images into two classes - natural and anthropogenic microhabitats. These are benchmarked against the results obtained using Manual classification.(Images © Copyright AD, RT 2022).}
        \label{fig:overview}
\end{figure}

We studied images of insects acquired from the Atlas of Living Australia (ALA) (\url{ala.org.au}), a database of Australian biodiversity, to determine the kinds of microhabitat in which insects were observed and photographed. The process is first outlined in point form and then described in detail in the subsections below and Fig. \ref{fig:overview}.
\begin{enumerate}

\item We extracted from ALA images of 3 insect species that inhabit a variety of Australian urban and natural environments to create a dataset.

\item A pre-processing step was performed to automatically segment out the pixels containing insects from each image in our dataset, leaving a derivative dataset with only the insects' backgrounds remaining in the images.

\item We developed, documented and refined a manual, repeatable process for classifying the insect image backgrounds as containing either natural or anthropogenic microhabitat.

\item We trained and tested convolutional neural networks to extract features from insect image backgrounds that would enable their classification into natural and anthropogenic microhabitats.

\item We developed a Support Vector Machine (SVM) to automatically classify features of the insect image backgrounds into classes corresponding to natural or anthropogenic microhabitats and benchmarked this classification against the manually determined results.
\end{enumerate}
We tested the hypothesis that an analysis of insect image backgrounds would reveal variation in the extent to which insect species occupy natural and anthropogenic microhabitats when they are observed. We explored this hypothesis for European honey bees (\textit{Apis mellifera}), European wasps (\textit{Vespula germanica}), and drone flies (\textit{Eristalis tenax}). These species were selected as they are frequently observed in Australian urban and natural environments, and are known to have a variety of foraging preferences likely to result in diverse but observable microhabitat utilisation. Intuitively we expected that insects that forage from floral resources would be documented utilising natural microhabitats more often than scavengers and predatory insects, but we were unsure of the extent to which this would be true among the studied species, and whether or not significant differences would appear in their photographic documentation. We tested this hypothesis using two different methods:

\textbf{Manual microhabitat identification}: We developed and applied a manual, repeatable method or algorithm to classify insect image backgrounds.

\textbf{Automated microhabitat identification}: We developed and applied a machine learning tool to classify insect image backgrounds automatically. This was benchmarked against the manual method.

In the following sections we outline our methods in detail.

\subsection{Insect image retrieval}
\label{ALA}

We used ALA as our source of ecological entomological data since the data are often collected by experts and have some chance of being evaluated for quality (\url{https://www.ala.org.au/data-quality-project/}). ALA contains occurrence records with text and often image data. The platform offers an Application Programming Interface (API) that makes the process of searching and downloading data for specific species and within specific geographic areas straightforward. We used this API to extract taxon name, species class, land-use type, species id, observation latitude and longitude, identification verification status and image URLs for each target insect species. Image URLs for each record were stored in a file with the other data and later downloaded using a Python script within the following constraints.

We constrained the software to download only images within the geographic region from $-10.41$ to $-43.38$ degrees latitude and $113.78$ to $153.45$ degrees longitude which comprises of the whole Australian mainland. Any records missing latitudinal or longitudinal data were ignored. We have also restricted the software to download images which have their identification verification status as 'research' and 'confident' to ensure the images that we used are taxonomically confirmed and identified by experts. Records with other status or no status were discarded. Our ecologist co-author visually "sanity checked" images to further ensure they were of the target species.

We downloaded image data for three insect species from ALA, honey bees (\textit{Apis mellifera}), drone flies (\textit{Eristalis tenax}) and European wasps (\textit{Vespula germanica}).
Honey bees were chosen as they are a recognisable, abundant, and important invasive pollinator insect in Australia \citep{https://doi.org/10.1111/icad.12178}. They are also relatively slow moving insects that visit flowers, two traits that make them a popular subject for amateur photographers. Drone flies also act as pollinators \citep{Howlett_Gee_2019} and, like bees, are commonly found in Australian urban gardens and natural landscapes on and around flowers and natural vegetation. European wasps are an invasive introduced species in Australia that is particularly abundant in eastern and southern regions \citep{spradbery1992distribution}. They are a declared and prominent pest in Australian urban areas effecting people and animals \citep{cook2019quantifying}. These wasps are, intuitively speaking, often spotted near general waste, food and food waste as they are attracted by sugars and meat. Due to their salient yellow and black patterning and their proximity to humans in urban areas, wasps have also become a subject of photography (and are sometimes mistaken for bees). European wasp's differences in behaviour from bees and flies makes them a potentially interesting contrast to pollinators. They provide an important opportunity for this study to test the clarity with which insect microhabitat occupation is evident in the image records of common species.

\begin{figure}[H]
     \centering
     \includegraphics[{width=15cm}]{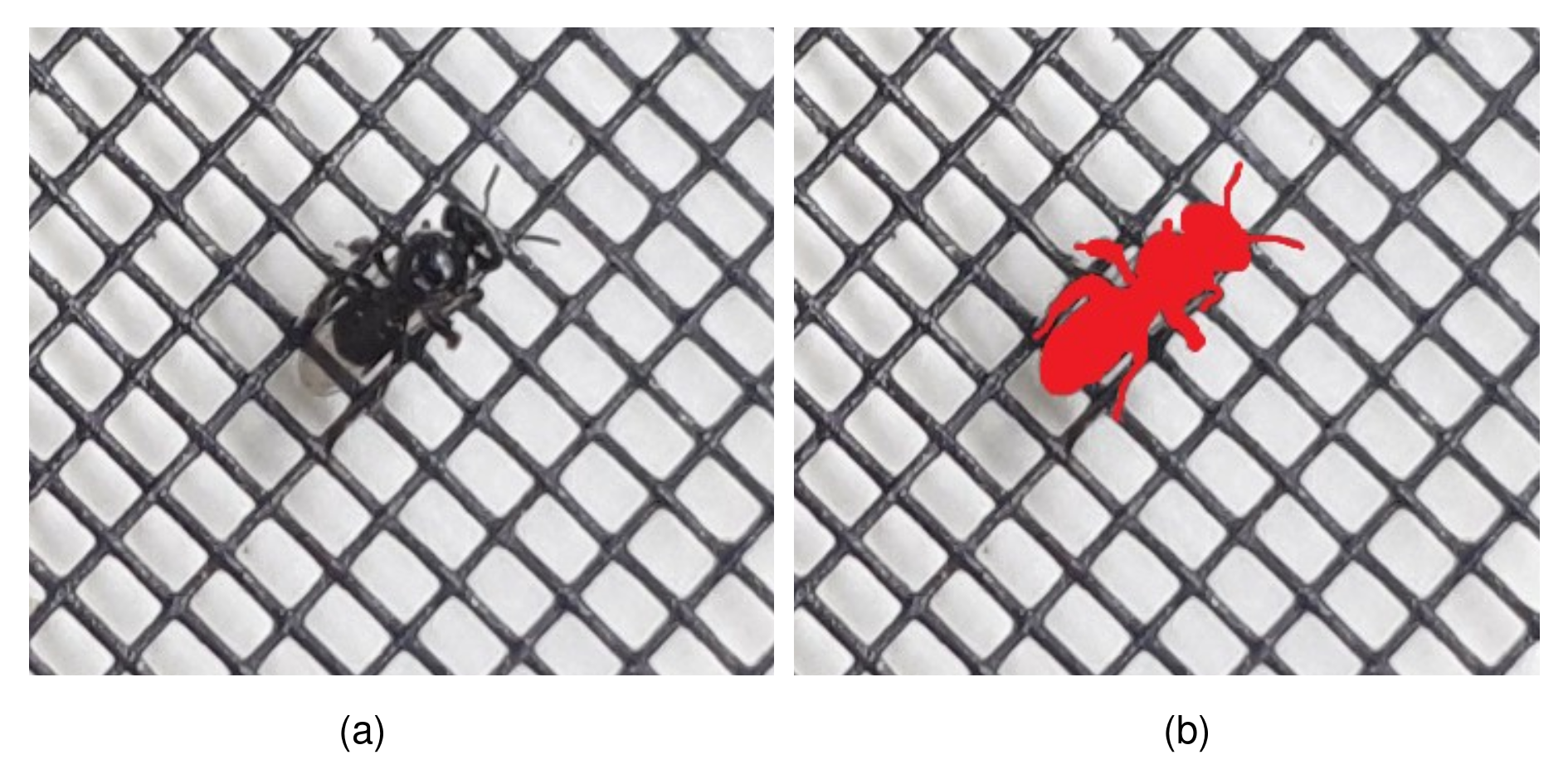}
        \caption{Example insect segmentation using YOLACT. (a) Raw image. (b) The insect is segmented out from the image to obtain the background. This provides evidence of the insect's microhabitat at the time of observation in a form that minimises the chance that a classifier might learn to identify the insect species and then infer the microhabitat by association. (Removed pixels have been highlighted in red for illustrative purposes.) Image © Copyright RT 2022.}
        \label{fig:beesegmentation}
\end{figure}

\subsection{Image insect subject removal}
\label{segmentation}
In this pre-processing step we segment out insect pixels from images to isolate image background pixels from which to classify observations of insect microhabitats. Failure to conduct this step may result in the classifier biasing its background classification based on its ``knowledge'' of the insect species contained within the image. Hence it plays an important role in ensuring any conclusions are meaningful. For the insect removal step we used the real-time instance segmentation algorithm, YOLACT (\textbf{Y}ou \textbf{O}nly \textbf{L}ook \textbf{A}t \textbf{C}oefficien\textbf{T'}s) \citep{bolya2019yolact}, pre-trained with the MS COCO dataset \citep{lin2014microsoft}. This is a large-scale object-detection and segmentation dataset of object classes including humans, animals, planes, ships, etc., but not insects – hence, YOLACT in its basic form was unable to directly segment insects from our dataset. We therefore fine-tuned the pre-trained model with 200 labelled ALA images of European honey bees and 200 images of European wasps. Fifty images of these classes were taken from the remaining dataset for fine-tuning the classifier. None of the images used in training were reused for experimental processes. The training and validation was performed on an NVIDIA P100 GPU. Due to the visual similarity between all insects in our study, we were able to successfully segment out their bodies using this trained model (e.g. Fig. \ref{fig:beesegmentation}).

\begin{figure}[H]
     \centering
\centerline{\includegraphics[width=17.5cm]{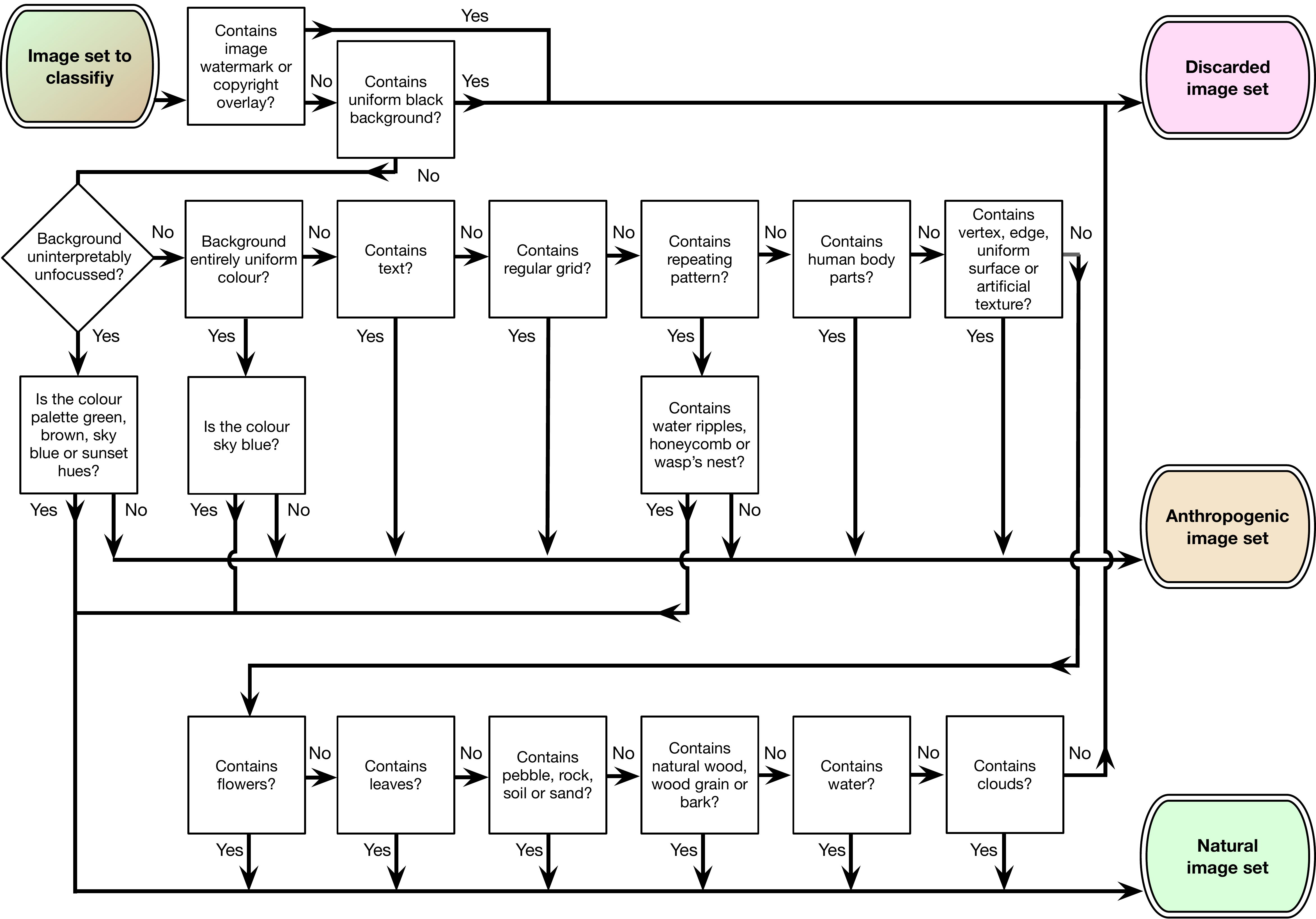}}
        \caption{Manual image background classification algorithm based on human observation to create training and validation datasets to train our software classification algorithm.}
        \label{fig:humanclsfr}
\end{figure}

\subsection{A manual algorithm to classify insect image backgrounds}
\label{method:manual}
We created an algorithm based on visual (human) observations of ALA image backgrounds (Fig. \ref{fig:humanclsfr}). We manually selected a subset of 500 images of \textit{Apis mellifera} (honey bees) with a variety of backgrounds to refine a human classification procedure. This process was necessary for us to articulate our understanding of what constitutes anthropogenic and natural microhabitats - a distinction without which training any software classifier would be at best problematic, and at worst nonsensical. Once this process was formally documented (Fig. \ref{fig:humanclsfr}) it was methodically followed to classify all images of the three insect species, drone flies, honey bees and European wasps.

\subsection{A software algorithm to classify insect image backgrounds}
\label{train_model}
\begin{figure}[H]
     \centering
     \includegraphics[{width=18.5cm}]{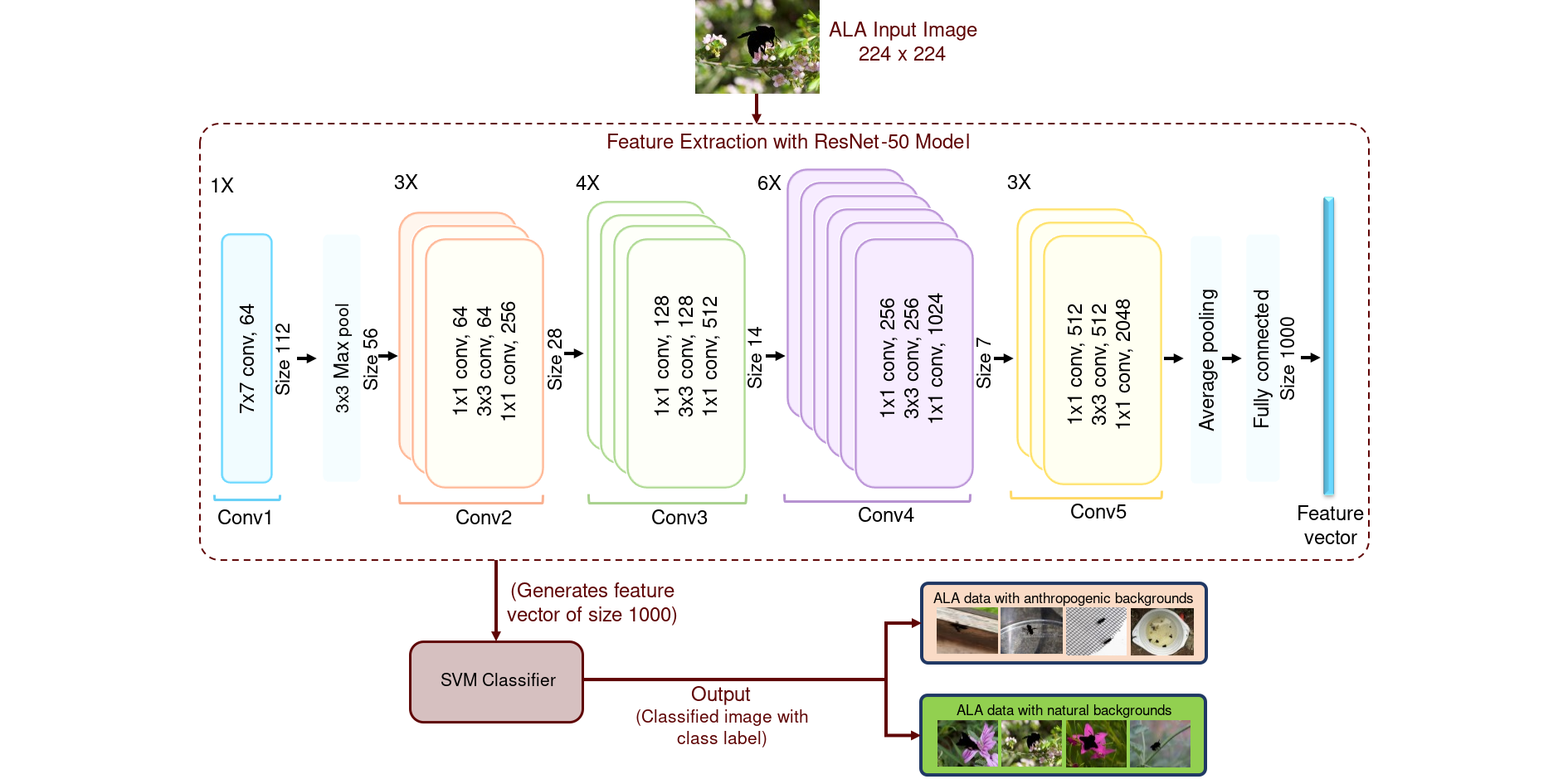}
        \caption{The software classifier showing the structure of ResNet-50 generating a feature vector of size 1000. This is used to train the SVM classifier to classify the insect image backgrounds into two classes. \textit{The notation `$axa$, b' in the convolution blocks 1-5 represent a filter of size a and b channels. The output of each convolution block is represented by `Size x'. The repetition of each square shape in each convolution block represents the repetition of each unit.}  }
        \label{fig:ResNet50}
\end{figure}

In this step we develop a machine learning classifier to distinguish the two classes of image background that broadly describe insect microhabitat as natural or anthropogenic (Fig. \ref{fig:ResNet50}).

\subsubsection{Creating a labelled dataset of image backgrounds for training}
\label{method:manual2}
To create image sets with which to train and test a software microhabitat classifier, we used the manual algorithm discussed in Section \ref{method:manual} and classified 500 images of European honeybee and European wasp backgrounds into equal sets of 250 natural and 250 anthropogenic microhabitats. These, manually but methodically labelled, images constituted the training data for the software classifier described in Section \ref{feature_ext}. A separate set of 100 \textit{Apis mellifera} images labelled using the same methodical process was set aside to test the trained software classifier.

\subsubsection{Feature extraction of image background}
\label{feature_ext}
The next step was to gather standard visual information and low-level image features from the datasets that include traits such as colour, texture, shape, etc. \citep{shih2001intelligent}. These features can be used by classifiers or deep learning models for image segmentation, image classification and object detection. There are various techniques for conducting image feature extraction. With advancements in machine learning, specifically deep learning and convolution neural networks (CNNs), feature extraction from images can be performed automatically to obtain high levels of classification accuracy \citep{acharya2020real}. Therefore, we used a deep-learning model to extract features of our image backgrounds for training a classifier model. We adopted a ResNet-50 model \citep{he2016deep} for extracting image background features for classification.  ResNets (Residual Networks) are easy to train with reduced complexity, even though they have deeper layers than CNN models, because of the presence of skip connections between the input and output of each block \citep{he2016deep}.  ResNet has different variants with a variety of convolution layers. We trialled three depths, ResNet-18, -50 and -101, and compared their training times and validation accuracies. They behaved similarly but ResNet-50 had a marginally higher validation accuracy (Section \ref{feature-ext})leading us to choose it for feature extraction.

ResNet-50 is made up of five convolutional blocks stacked on top of one another (Fig. \ref{fig:ResNet50}). The image features are extracted from the fully connected layers pre-trained with the ImageNet database (\url{https://image-net.org/index.php}). The input to the network is an image of 224x224 pixels. Features extracted from the deeper layers  class-specific properties such as shape, texture and colour, and hence provide a better classification performance than features extracted from the shallower layers \citep{10.1007/978-3-319-10590-1_53}. Therefore, we extracted features from the last layer, a fully connected layer, that outputs a 1000-dimensional feature vector.

\subsubsection{Training and validating the classifier model}
\label{training}
Deep learning is increasingly used for image classification \citep{JENA2021104803, 10.1007/978-981-16-2934-1_9, LI2021101460}. The literature reports research where purpose-built datasets are used to train species-specific image-classifiers for identifying insects, such as crop pests, in images \citep{9825681}. However, training such models for specific purposes requires relatively large datasets, it is time-consuming and computationally expensive. 
To overcome these shortcomings, we used Support Vector Machines (SVMs) \citep{cortes1995support} as our classifier model. SVMs are machine learning models that transform non-linear separable problems to linearly separable problems. They have a high generalising ability compared to other classifier models \citep{CERVANTES2020189} and are capable of delivering high classification accuracy \citep{durgesh2010data} with small datasets and little time and computational expense. We used the extracted background features generated in our previous step to train an SVM model with 5-fold cross validation. The training and validation of our classifier was run on a basic laptop using its CPU (Intel(R) Core(TM) i7-9850H, CPU clock speed=2.60 GHz).

\section{Results}
\label{results}
Section \ref{manual-class} presents the results of our manual classification algorithm. Sections \ref{feature-ext} and \ref{res:classifier_model} provide training and validation results for our automatic classifier. Section \ref{case-study-B} provides the results of our insect microhabitat studies.

\subsection{The discovery of exceptions and special cases during the manual classification of image backgrounds}
\label{manual-class}
\begin{figure}[H]
    \centering
    \setkeys{Gin}{width=\linewidth}
     \begin{subfigure}[b]{0.4\textwidth}
         \centering
         \includegraphics[width=\textwidth, height=5.5cm]{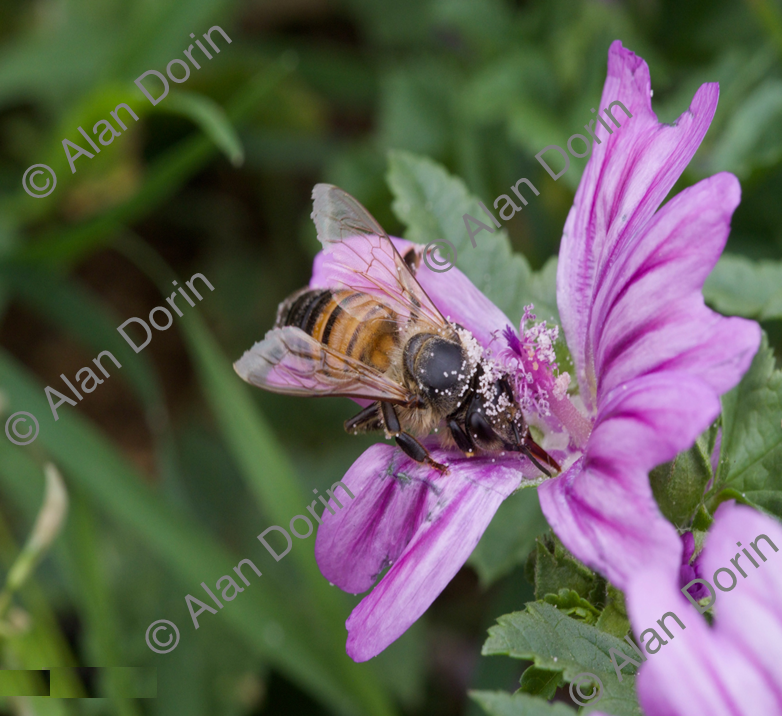}
         \caption{}
         \label{fig:exception_4}
     \end{subfigure}
     \begin{subfigure}[b]{0.4\textwidth}
         \centering
         \includegraphics[width=\textwidth, height=5.5cm]{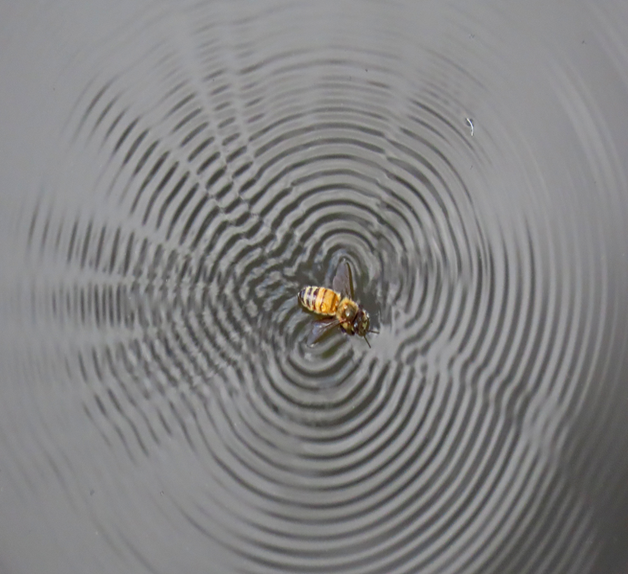}
         \caption{}
         \label{fig:exception_2}
     \end{subfigure}
     
     \begin{subfigure}[b]{0.4\textwidth}
         \centering
         \includegraphics[width=\textwidth, height=5.5cm]{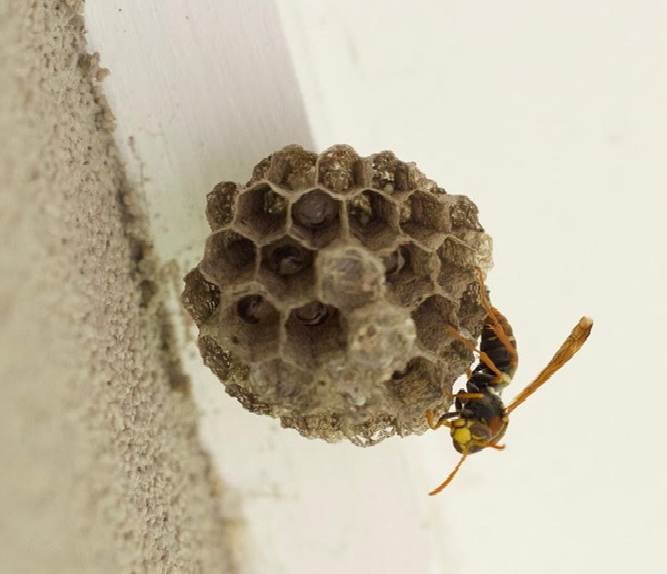}
         \caption{}
         \label{fig:exception_3}
     \end{subfigure}
     \begin{subfigure}[b]{0.4\textwidth}
         \centering
         \includegraphics[width=\textwidth, height=5.5cm]{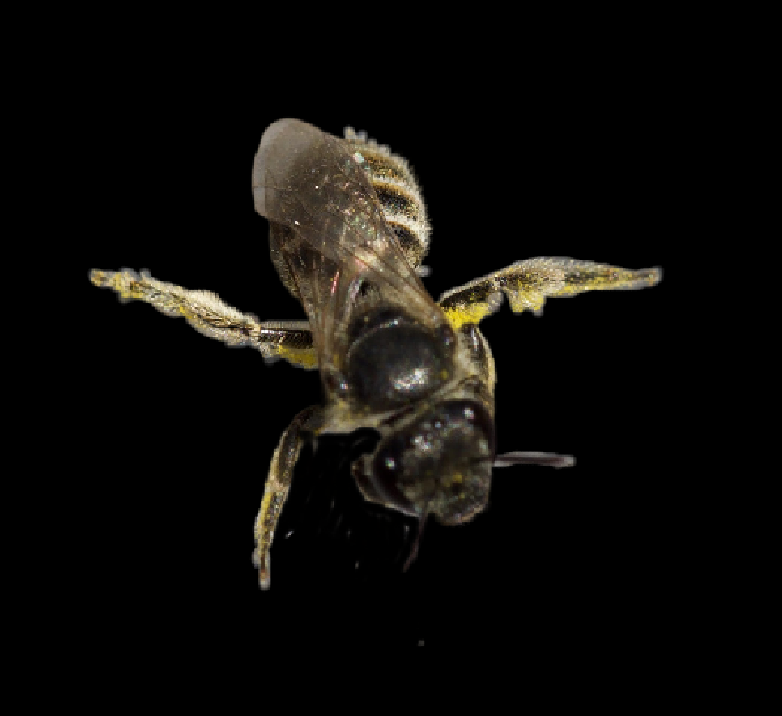}
         \caption{}
         \label{fig:exception_4}
     \end{subfigure}
        \caption{The kinds of exceptions encountered during manual microhabitat classification. (a) Watermarked images were discarded; (b) Regular water-ripples were classified as natural; (c) Insect nest structures were classified as natural even if the nest was attached to artificial substrate; (d) Images with black or removed backgrounds were discarded. Image (b) \url{https://images.ala.org.au/image/f2e756f0-1e7e-4436-855c-e23d8b00b643} © deborod 2020; licensed under CC-BY-NC 4.0; (a),(c),(d) © AD 2022.}
        \label{fig:exceptions}
\end{figure}

While manually classifying image backgrounds we encountered some exceptions and edge cases that lead to difficulties in labelling microhabitat types (Fig. \ref{fig:exceptions}). One edge case related to the presence of text in an image (Fig. \ref{fig:exception_4}). If text sits within the non-insect pixels of an image background, such as on a handwritten paper note, or printed words on a page, packaging or signage, we would consider the insect to have been observed in an anthropogenic scenario. However, watermarks and copyright claims were also digitally superimposed on some images. Such artefacts of the image-making process (obviously) shouldn't be considered components of a depicted insect's microhabitat. To simplify the process of correctly determining microhabitat, we subsequently excluded images containing overlaid text from our study.
Regular grids or other repeated patterns present in image backgrounds are usually classified as anthropogenic according to our manual algorithm. However, exceptions to this rule were identified. For instance, we discovered water ripples forming repeated patterns and classified these as natural (Fig. \ref{fig:exception_2}). Similarly, insect nests, such as honey bee honeycomb structures or some wasps nests, may be grid-like or exhibit other repeated patterns. These may be either in anthropogenic or natural settings depending on the location (e.g. within a box-beehive, attached to a tree branch, or under a building's eaves Fig. \ref{fig:exception_3}). As our dataset has only a few insect nest images, we ignored the sub-classification of nests by location and classified them all as natural, even though typically, regular grid-like patterns are indeed human artefacts. We also discarded all images with pure black backgrounds (Fig. \ref{fig:exception_4}) as the lack of features made it impossible to classify them.

\subsubsection{Results of feature extraction from image backgrounds}
\label{feature-ext}
\begin{figure}[H]
     \centering
     \includegraphics[{width=15cm}]{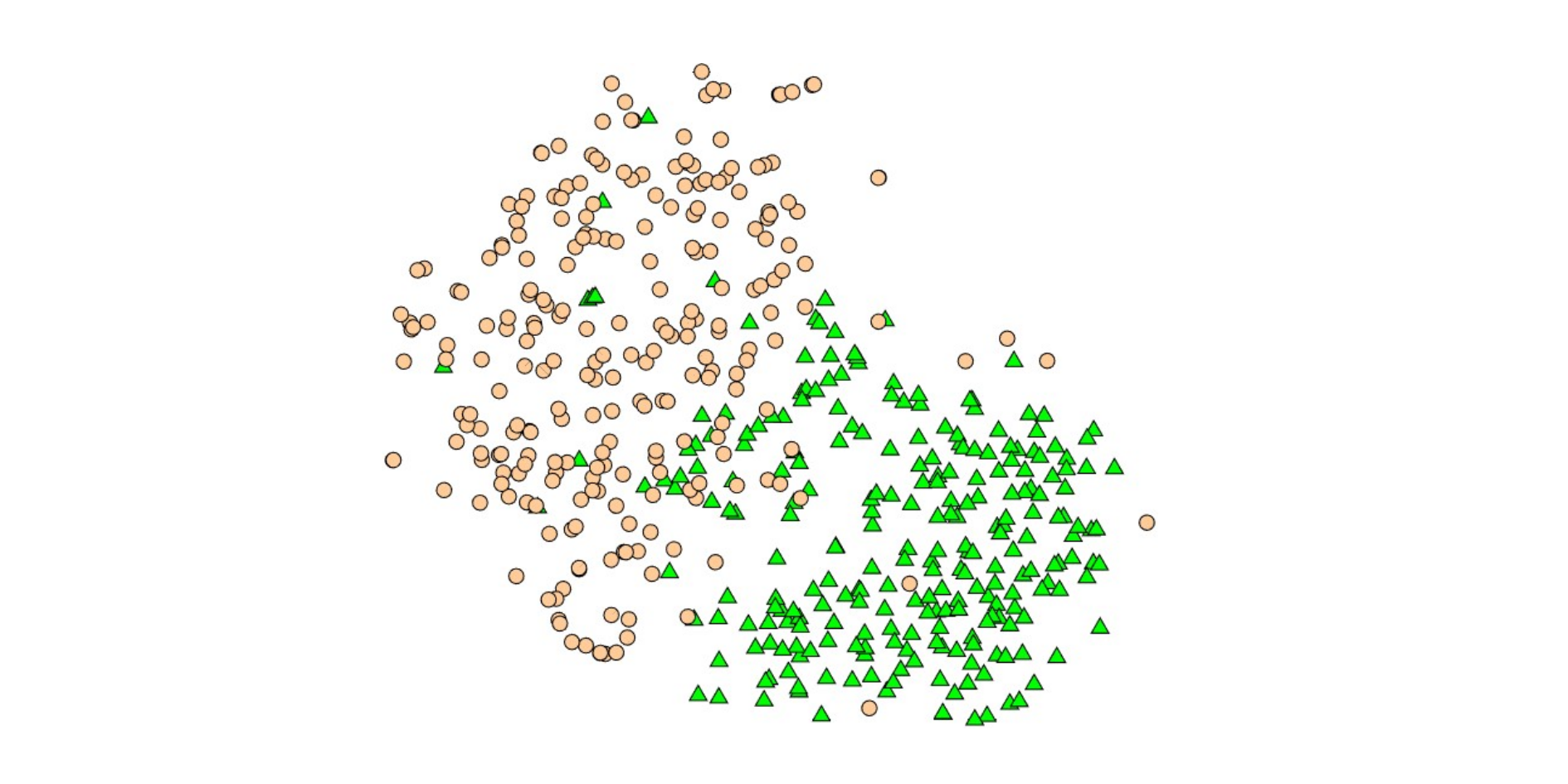}
        \caption{Visualisation of extracted features using a pre-trained ResNet-50 CNN model showing that the background microhabitat classes are well structured. \textbf{Green triangles} represent `natural' microhabitats and the \textbf{light brown circles} represent `anthropogenic' mirohabitats. }
        \label{fig:t-SNE}
\end{figure}

We tested the performance of pre-trained ResNet-18/50/101 models by training each with the same 500 honey bee images. This dataset contained 250 images of natural backgrounds and 250 anthropogenic backgrounds as determined by the manual algorithm. 
The training times for all three models were similar, with ResNet-50 being marginally faster. A pre-trained ResNet-50 CNN model was therefore selected for our study, and the last layer before its classification layer was used to extract 1000 features for training our classifier model. These features are visualised in Fig. \ref{fig:t-SNE} using a t-SNE algorithm \citep{van2008visualizing}. This is a non-linear dimensionality reduction method for visualising high-dimensional data by providing each data-point a location in 2-D or 3-D space. In this case it shows that our two classes, natural and anthropogenic microhabitats, are well structured.

\begin{table}[ht]
\centering
\begin{tabular}{|l|l|l|l|}
\hline
 & \textbf{ResNet-18} & \textbf{ResNet-50} & \textbf{ResNet-101} \\
\hline
\textbf{Training time} & 1.25 s & 1.2 s & 1.22 s \\
\hline
\end{tabular}
\caption{\label{tab:model_comp}Comparison of training time for pre-trained ResNet-18/50/101 models for extracting features of backgrounds of 500 honey bee images.}
\end{table}

\subsubsection{Results of classifier model training and validation}
\label{res:classifier_model}
After the SVM classifier was trained with image features (Section \ref{feature_ext}) extracted from 500 honey bee images (training time = ~30 secs, on laptop CPU), 88 random images were selected from the remaining honey bee images to test the model's classification performance (test time = 0.25 seconds on laptop CPU). We performed 5-fold cross validation for our SVM model. We compared the validation inaccuracies of the SVM classifier model using the features extracted from ResNet-18/50/101 models (Table \ref{tab:model_comp_2}). We found that features extracted from ResNet-50 generated marginally better validation accuracy than ResNet-18 or -101 and obtained a validation accuracy for training the SVM of $96.4\%$ and a test accuracy of $97.4\%$. This strengthens our motivation for using ResNet-50 as our image feature extractor. Some classification results obtained from the trained SVM model did not match their manual classifications, i.e. the method's accuracy was not $100\%$. (Section \ref{case-study-B})

\begin{table}[ht]
\centering
\begin{tabular}{|l|l|l|l|}
\hline
 & \textbf{ResNet-18} & \textbf{ResNet-50} & \textbf{ResNet-101} \\
\hline
\textbf{Validation accuracy} & $95\%$ & $96.4\%$ & $94\%$ \\
\hline
\end{tabular}
\caption{\label{tab:model_comp_2}Comparison of the validation accuracy of the SVM classifier using the features of backgrounds of 500 honey bee images extracted using ResNet-18/50/101 models  }
\end{table}

\subsubsection{Results of comparison between manual and automated microhabitat identification methods}
\label{case-study-B}
\begin{figure}[H]
     \centering
     \includegraphics[{width=18.5cm}]{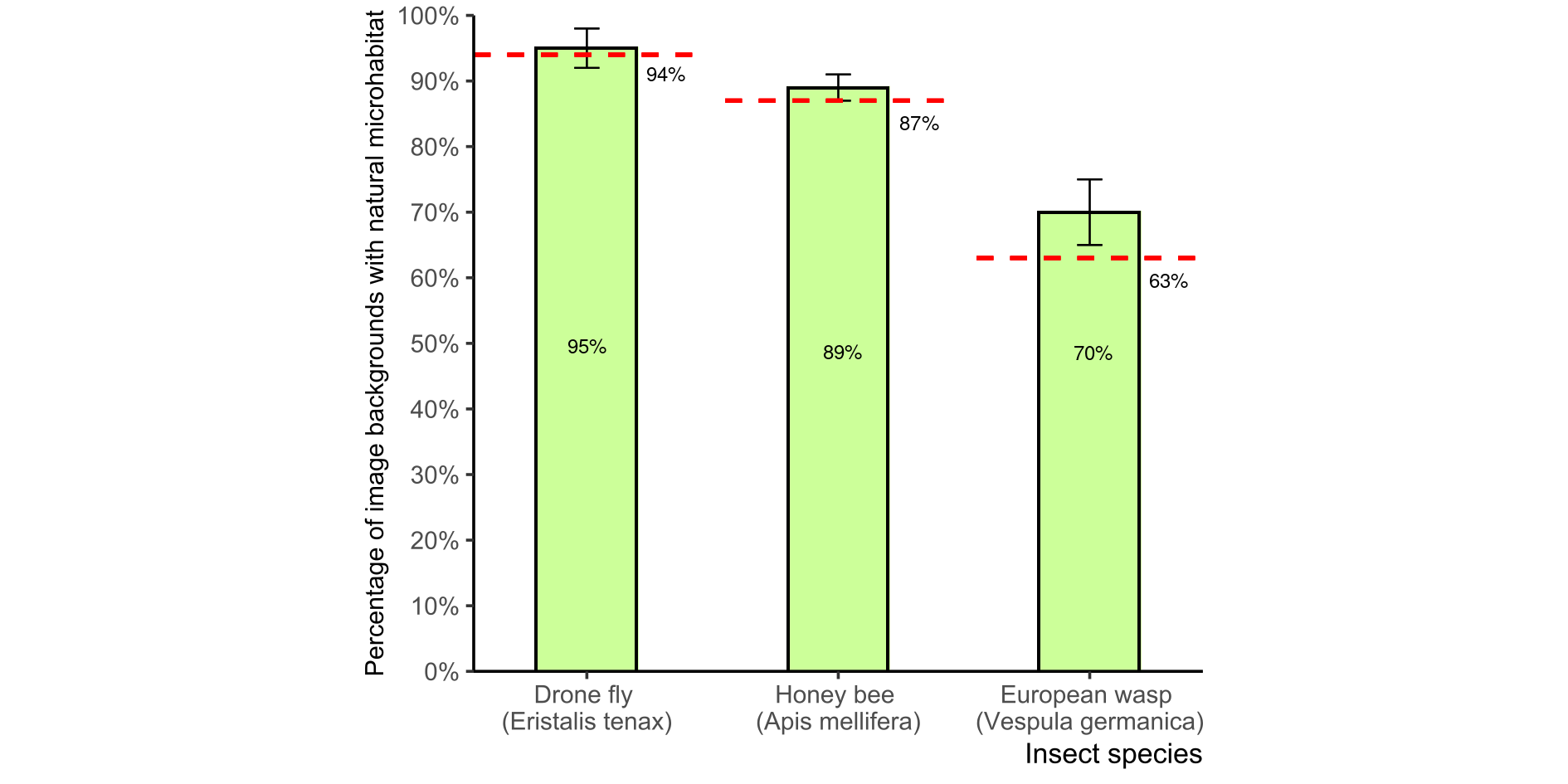}
    
        \caption{Plot showing the percentage of ALA target insect species images with `natural' backgrounds calculated using the manual classification method (solid green bars) and the automated classifier measurement of the same value (red dashed lines). Error bars represent $95\%$ confidence intervals calculated using the Wilson score interval test \citep{wilson1927probable}. Manual/automated method results: Drone flies = $95\pm3\%$ / $94\%$ Honey bees = $89\pm2\%$ / $87\%$, European wasps = $70\pm6\%$ / $63\%$.}
        \label{fig:ALA-comp}
\end{figure}

Following our download of ALA's images of honey bees (7654 images), European wasps (1026 images) and drone flies (706 images) within the extent of mainland Australia and manual removal of irrelevant images, over 9000 images of the target species remained (Table \ref{tab:ALA-img-count}). The classification of these images' backgrounds using both manual and automated methods into natural and anthropogenic microhabitats is provided (Fig. \ref{fig:ALA-comp}).

\begin{table}[h]
\centering
\begin{tabular}{|c|c|c|c|}
\hline
 \textbf{Common name} & \textbf{Search term (scientific name)} & \textbf{Number of images returned} \\
\hline
\textbf{honey bee} & \textit{Apis mellifera} & 7669 \\
\hline
\textbf{european wasp} & \textit{Vespula germanica}  & 1026 \\
\hline
\textbf{drone fly} & \textit{Eristalis tenax}  & 706 \\
\hline
\end{tabular}
\caption{\label{tab:ALA-img-count}ALA images collected by searching for species' scientific names on 21\textsuperscript{st} Nov 2022 and ascertained by superficial visual inspection to be images of the target species.}
\end{table}

The results from the automated method's classification of natural (versus anthropogenic) backgrounds is slightly, but consistently, lower than that produced by the manual algorithm across all species studied. This is discussed in Section \ref{discussion}. Irrespective of the classification algorithm, the drone flies in our data were photographed almost entirely in natural microhabitat, honey bees only slightly less so. By contrast, European wasps were photographed much more prominently in anthropogenic microhabitats, even though the majority of their images did still show natural microhabitats. 

In order to determine if these three species of insects are part of significantly different distribution and find the confidence interval of the population proportion `p' for each distribution, we have performed a Wilson score interval test \citep{wilson1927probable}. This test provides $95\%$ confidence values for the natural image background prediction using our manual method.Results show $95\pm3\%$ drone flies have natural backgrounds, honey bees and European wasps have $89\pm2\%$ and $70\pm6\%$ natural backgrounds respectively when classified manually.

\begin{figure}[H]
     \centering
     \includegraphics[{width=15cm}]{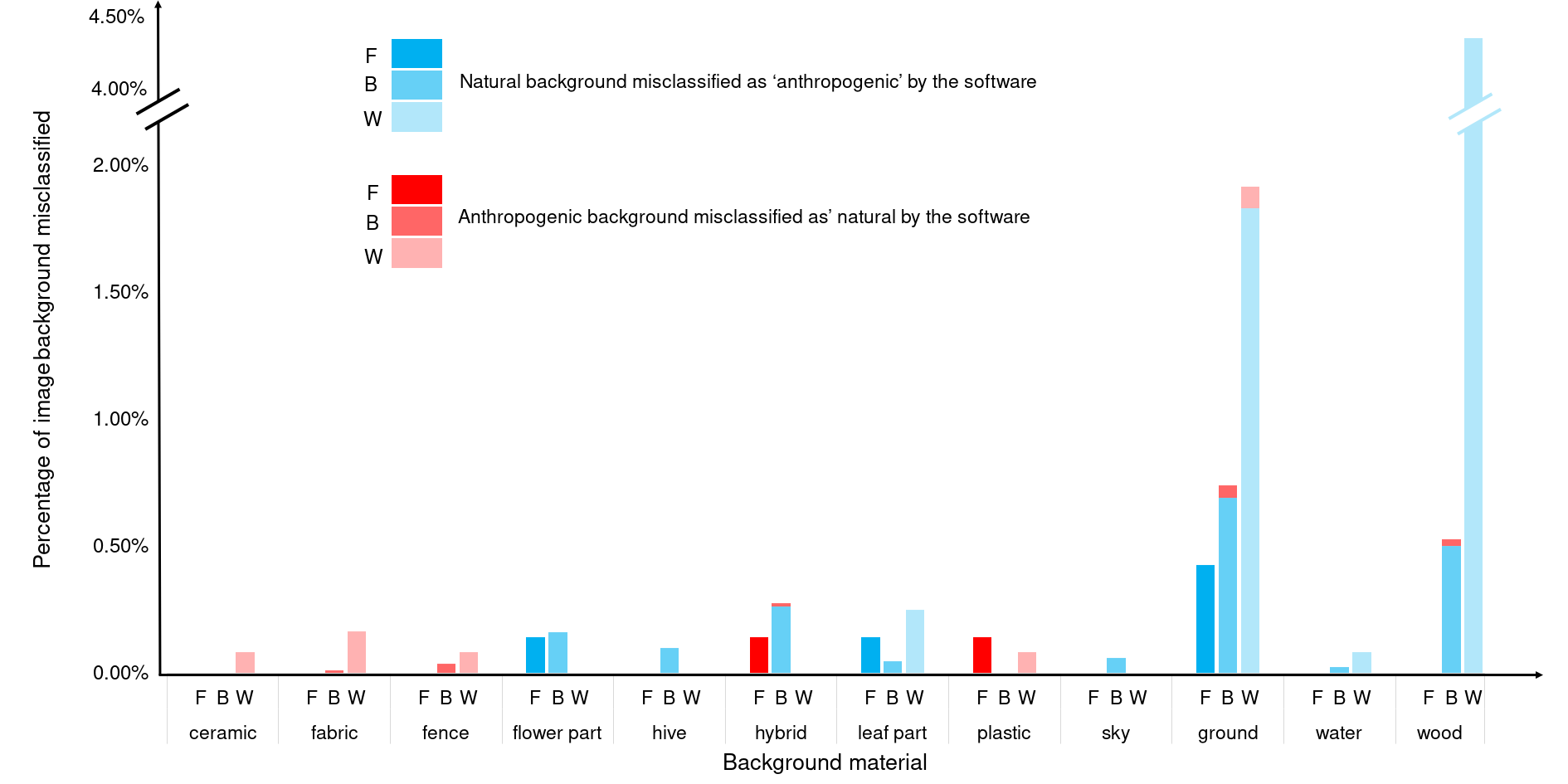}
         \label{fig:y equals x}
     \caption{Types of background mis-match between the automated software and the manual algorithm for each insect species. Blue shades represent natural backgrounds misclassified as anthropogenic by the automated model. Red shades represent anthropogenic backgrounds misclassified as natural by the automated model. `F', `B' and `W' indicate classification discrepancies for drone \textbf{F}lies, honey \textbf{B}ees and European \textbf{W}asps respectively. Detailed descriptions of classification mismatch types are given in Section \ref{case-study-B} and examples appear in Fig. \ref{fig:miscla_eg}.}
     
     \label{fig:misclassification}
\end{figure}

\begin{figure}[H]
     
     \begin{subfigure}[b]{0.159\textwidth}
         \centering
         \includegraphics[width=\textwidth, height=3cm, keepaspectratio]{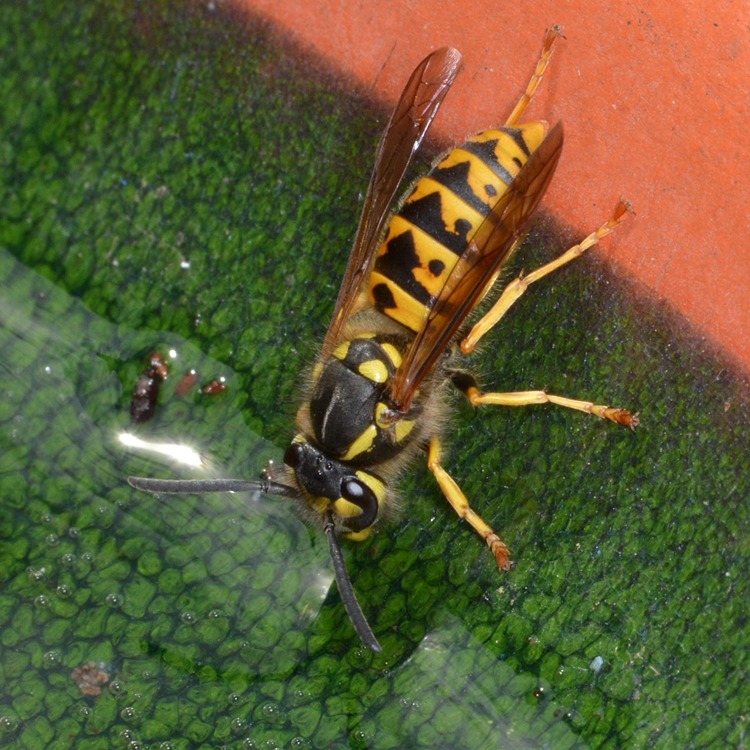}
         \caption{}
         \label{ceramic}
     \end{subfigure}
     \begin{subfigure}[b]{0.159\textwidth}
         \centering
         \includegraphics[width=\textwidth, height=3cm, keepaspectratio]{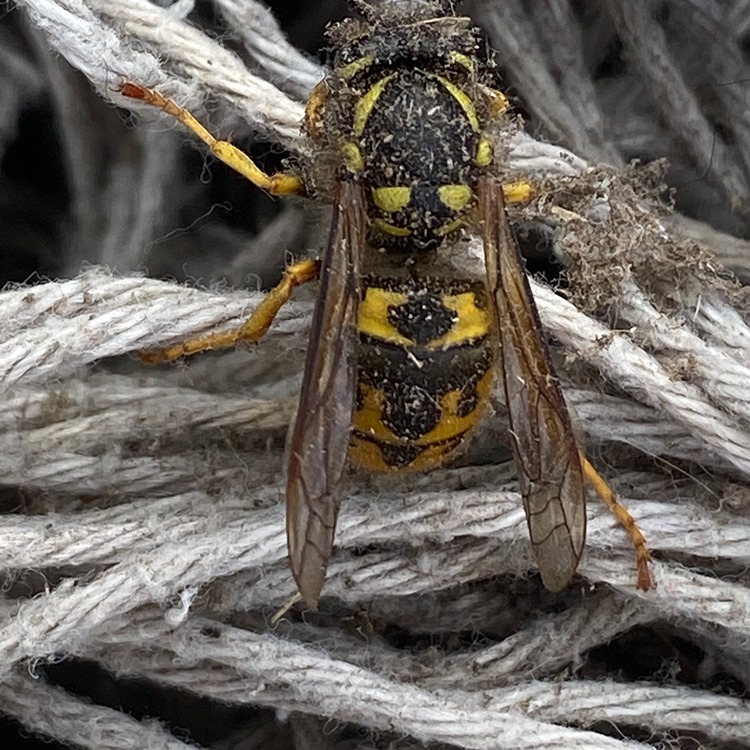}
         \caption{}
         \label{fabric}
     \end{subfigure}
          \begin{subfigure}[b]{0.159\textwidth}
         \centering
         \includegraphics[width=\textwidth, height=3cm, keepaspectratio]{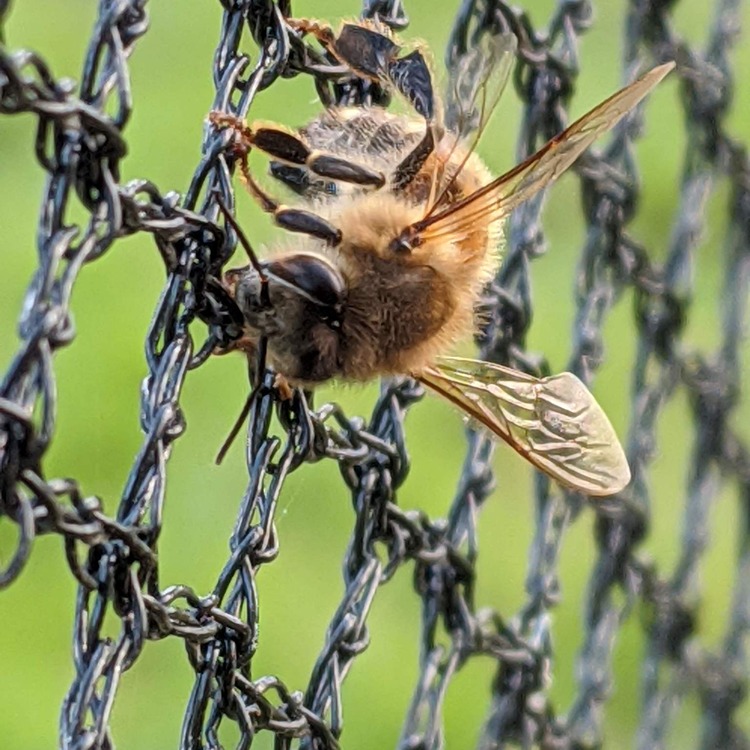}
         \caption{}
         \label{fence}
     \end{subfigure}
          \begin{subfigure}[b]{0.159\textwidth}
         \centering
         \includegraphics[width=\textwidth, height=3cm, keepaspectratio]{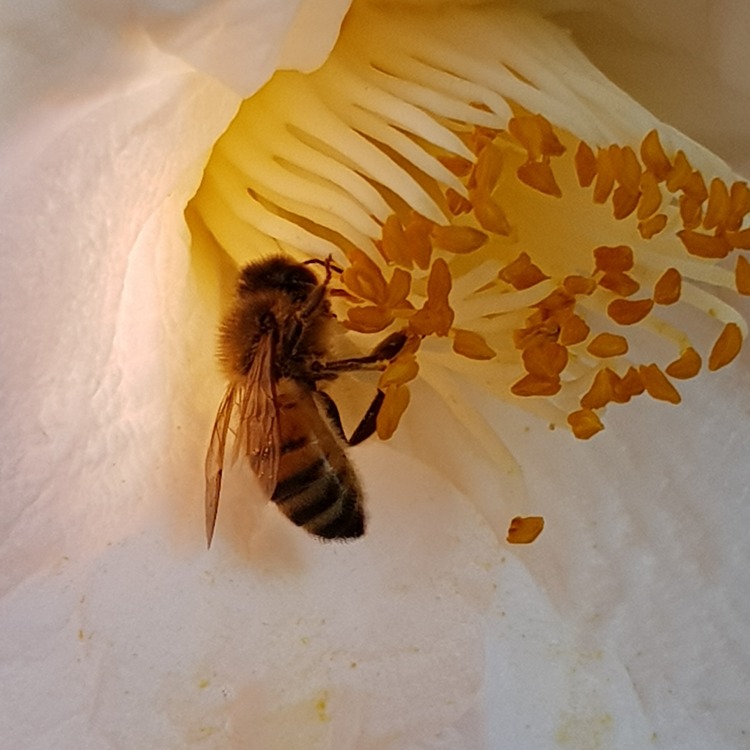}
         \caption{}
         \label{flower_part}
     \end{subfigure}
     \begin{subfigure}[b]{0.159\textwidth}
         \centering
         \includegraphics[width=\textwidth, height=3cm, keepaspectratio]{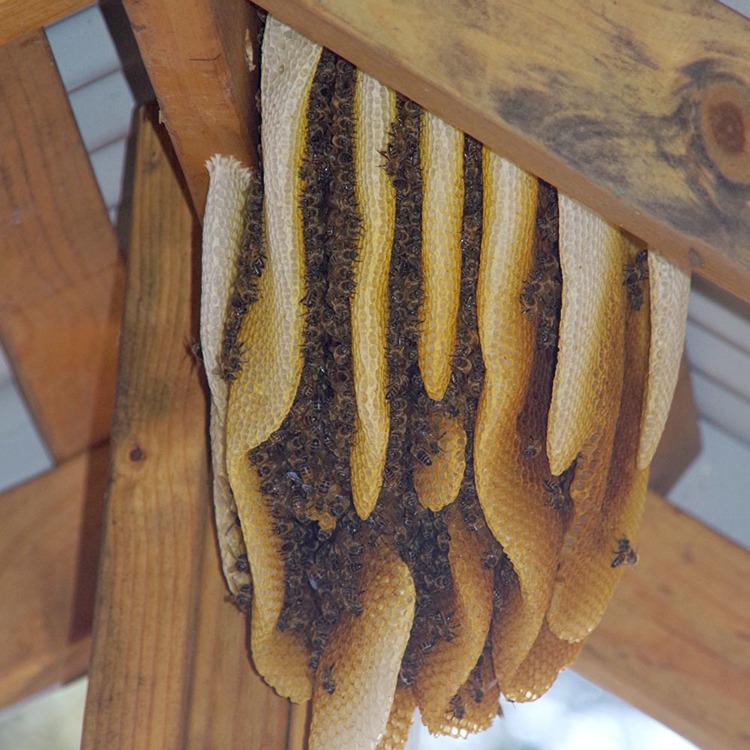}
         \caption{}
         \label{hive}
     \end{subfigure}
     \begin{subfigure}[b]{0.159\textwidth}
         \centering
         \includegraphics[width=\textwidth, height=3cm, keepaspectratio]{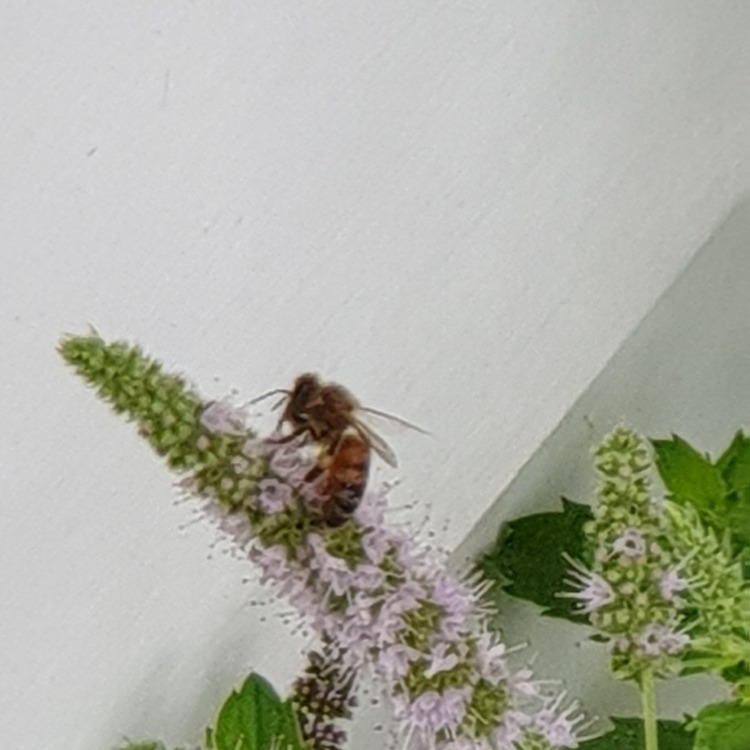}
         \caption{}
         \label{fig:hybrid}
     \end{subfigure}
     \newline
     \begin{subfigure}[b]{0.159\textwidth}
         \centering
         \includegraphics[width=\textwidth, height=3cm, keepaspectratio]{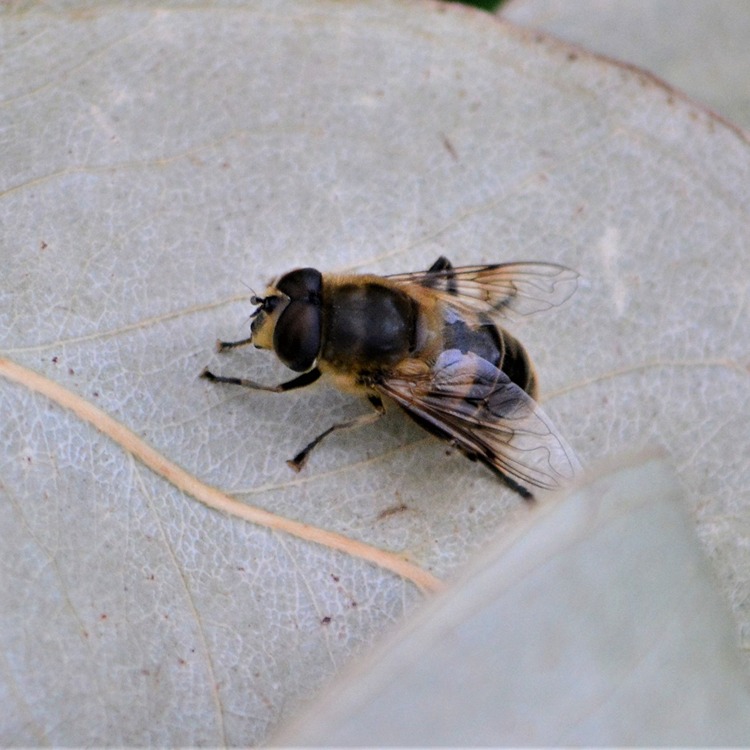}
         \caption{}
         \label{leaf_part}
     \end{subfigure}
     \begin{subfigure}[b]{0.159\textwidth}
         \centering
         \includegraphics[width=\textwidth, height=3cm, keepaspectratio]{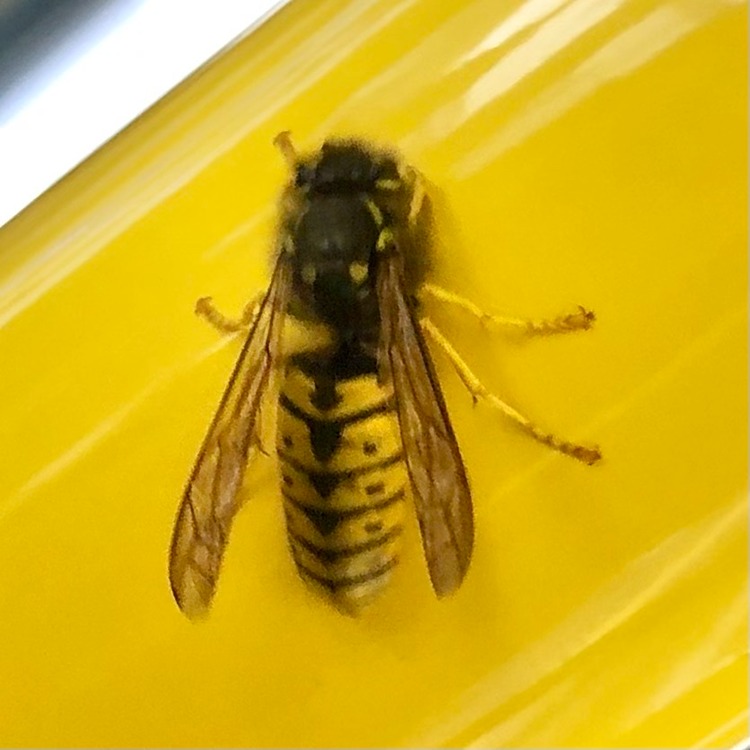}
         \caption{}
         \label{plastic}
     \end{subfigure}
          \begin{subfigure}[b]{0.159\textwidth}
         \centering
         \includegraphics[width=\textwidth, height=3cm, keepaspectratio]{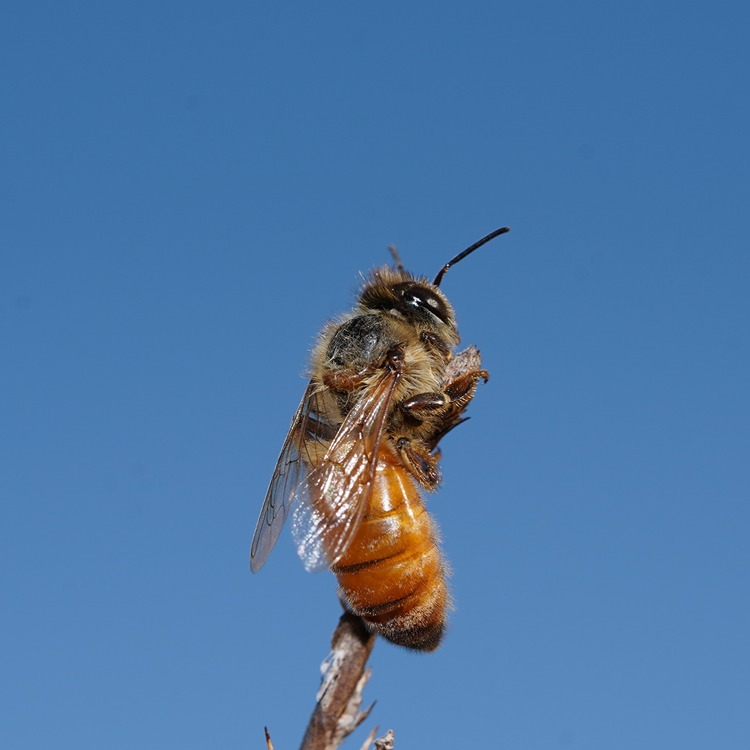}
         \caption{}
         \label{sky}
     \end{subfigure}
          \begin{subfigure}[b]{0.159\textwidth}
         \centering
         \includegraphics[width=\textwidth, height=3cm, keepaspectratio]{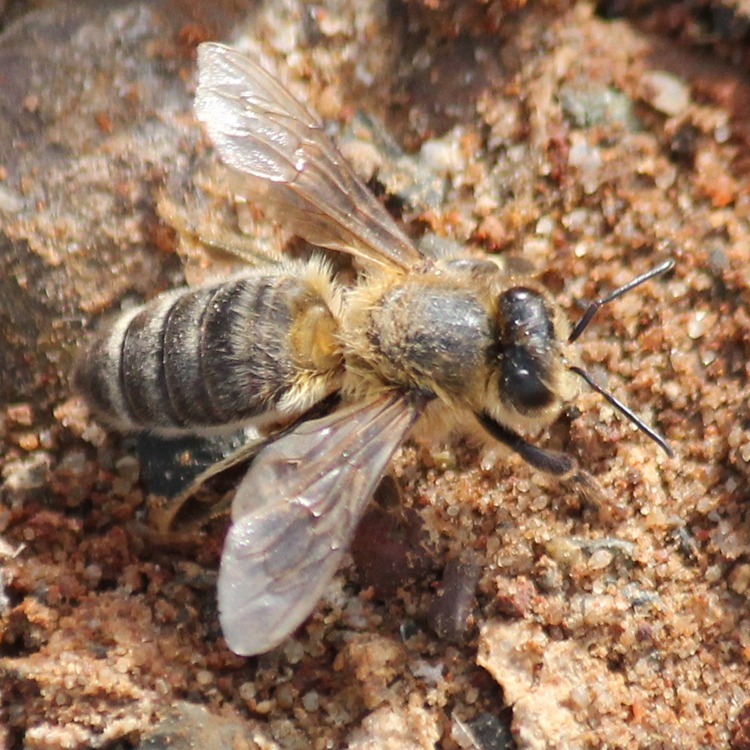}
         \caption{}
         \label{soil}
     \end{subfigure}
     \begin{subfigure}[b]{0.159\textwidth}
         \centering
         \includegraphics[width=\textwidth, height=3cm, keepaspectratio]{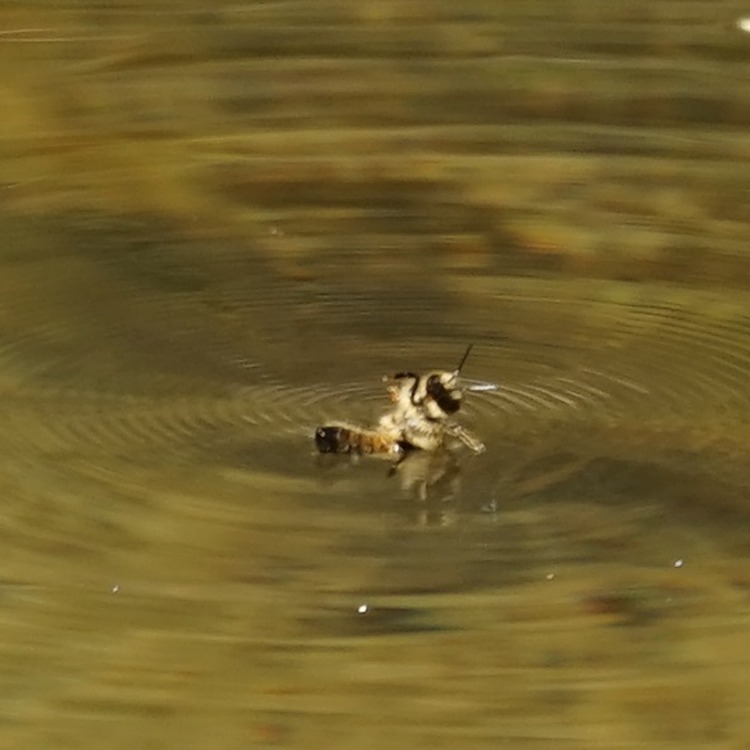}
         \caption{}
         \label{fig:water}
     \end{subfigure}
     \begin{subfigure}[b]{0.159\textwidth}
         \centering
         \includegraphics[width=\textwidth, height=2.8cm]{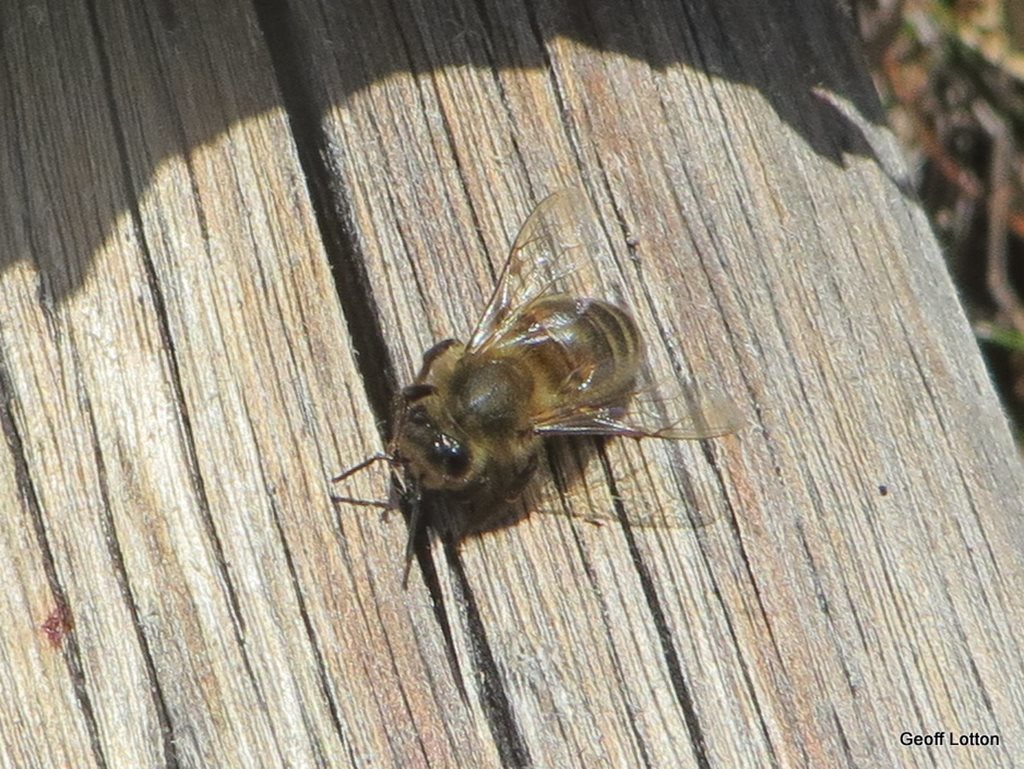}
         \caption{}
         \label{fig:wood}
     \end{subfigure}
        \caption{Examples of image background types occasionally mis-classified by the automated software: (a) ceramic; (b) fabric; (c) fence; (d) flower part; (e) hive; (f) hybrid; (g) leaf part; (h) plastic; (i) sky; (j) ground; (k) water; (l) wood. Images (a-g, j-k) used under license BY-NC 4.0, (h) under license BY-SA 4.0, (i,l) under license BY-NC-SA 4.0. Image source (a) \url{https://images.ala.org.au/image/67e833c5-e949-481f-9d39-90ec2d30d27b} © Menura 2021; (b) \url{https://images.ala.org.au/image/169abe6a-c292-472c-9228-020e3d73bb09} © Kymelen 2021; (c) \url{https://images.ala.org.au/image/b014ce62-0953-492e-a05f-4e29cd1bc09f} © Paula Rivera 2021; (d) \url{https://images.ala.org.au/image/689bf48f-3dbc-472c-9e0d-967fab1aa97c} © Wild Days Wildlife Shelter 2021; (e) \url{https://images.ala.org.au/image/4789cdba-dff3-4ffb-9bc9-0804560e5b2c} © Russell Cumming 2018; (f) \url{https://images.ala.org.au/image/5073019c-3e46-4daa-98e9-a9e7f7aae287} © QuestaGame 2018; (g) \url{https://images.ala.org.au/image/9c648e56-2de8-44db-ae91-5909ee779034} © Rosemary Kidd, 2020; (h) \url{https://images.ala.org.au/image/6b7525de-f44c-44c6-986e-5001713b4307} © James K. Douch 2022; (i) \url{https://images.ala.org.au/image/52cf6df3-609f-4b88-8efe-ee3583162798} © Reiner Richter 2019; (j) \url{https://images.ala.org.au/image/99c73250-7fd6-4628-914b-1f1ffdd19f49} © Reiner Richter 2017; (k) \url{https://images.ala.org.au/image/48736075-fd49-4b02-9deb-afc5072b4d19} © Tim 2020; (l) \url{https://images.ala.org.au/image/39ed3fda-c300-49c9-a7e7-37f373ab63bd} © Geofflot 2017.}
        \label{fig:miscla_eg}
\end{figure}

Fig. \ref{fig:misclassification} shows the results of comparing the software image background classifications to the manual method. For $10$ of the $12$ listed background types discrepencies were less than $0.25\%$. In the case of the wasps, natural ground and wood backgrounds were misclassified as anthropogenic: $2\%$ ground, $4.5\%$ for wood (see Section \ref{discussion}). Twelve mismatch types were identified in classifying the image backgrounds:
\begin{itemize}
\item Ceramic : a ceramic or glazed surface that is anthropogenic (Fig. \ref{ceramic}). 

\item Fabric : cloth or fabric that is anthropogenic (Fig. \ref{fabric}).

\item Fence : a garden or park fence, often made of metal, that is anthropogenic (Fig. \ref{fence}).

\item Flower part : a close-up image of a small flower part, such as a petal or flower anthers, that is natural (Fig. \ref{flower_part}).

\item Hive : an insect hive or nest, often with a repeated or grid-like pattern that is natural (Fig. \ref{hive}).

\item Hybrid : an image background containing both natural and anthropogenic components, such as a background showing a flower set against a wall (Fig. \ref{fig:hybrid}). In such cases the human considered the location where the insect was situated to be the best indicator of microhabitat use, but the software sometimes classifies in the opposite way to the human, taking a broader view of the image background contents in making its classification (since this was how it was trained).

\item Leaf part : a close-up image of a leaf or parts of a leaf that is natural (Fig. \ref{leaf_part}).

\item Plastic : a plastic surface such as a serving plate, drink cup or plastic bag that is anthropogenic (Fig. \ref{plastic}).

\item Sky : the blue sky is natural (Fig. \ref{sky}).

\item Ground : soil, mud, sand, gravel, etc. are natural (Fig. \ref{soil}).

\item Water : water in a natural setting, such as a pond (Fig. \ref{fig:water}).

\item Wood : tree trunks, split or natural logs, wooden tables and benches, etc. were often natural but classified by the software as anthropogenic (Fig. \ref{fig:wood}). In a couple of cases, the background was identified manually as an anthropogenic picnic table or benchtop but the model classified it as natural.
\end{itemize}

\section{Discussion}
\label{discussion}
Our results demonstrate how image backgrounds can be analysed manually and automatically to generate important data on the kinds of microhabitats in which humans encounter insects. This is a novel approach to insect microhabitat studies derived from biodiversity data that might otherwise go unanalysed. Images online provide a potentially huge source of information on species occurrence \citep{elqadi2017mapping}, but as shown here, we can go further than simply using these images as evidence of an individual insect's presence. These images allow us to learn how species occupy the environment when they are encountered by humans. Although we restricted this study to ALA images for quality control reasons, previous work has demonstrated that machine learning and computer vision tools can correctly identify some species from social media posts \citep{elqadi2017mapping, BURKE2022101598}. Consequently, in the future, our method could be applied to social media image data as ``Incidental Citizen Science'' to deduce insect microhabitat, especially in urban areas where social media posts are most abundant. Such information can assist in the control of invasive species like the European wasp (and the honey bee for that matter). But it also allows us to obtain data on insect utilisation of urban habitats that might inform biodiversity management of pollinating species like honey bees (these aren't often viewed through the lens of invasive species, even in Australia) and drone flies.
\newline

Analysis of the microhabitats of the study species shows the large extent to which drone flies and honey bees are encountered by humans against natural backgrounds (Fig. \ref{fig:miscla_eg}). We found this to be true irrespective of the classification method (flies $95\%$ and bees $89\%$ using manual classification, $94\%$ for flies and $87\%$ for bees using the automated classifier). These insects are important pollinators. Their tendency to loiter on, in and around natural microhabitat features such as plants, trees and flowers is strongly supported by the data we acquired through image analysis. This highlights also the need for the explicit provision of natural microhabitats within our urban built environments to cater for these insects \citep{10.18475/cjos.v52i2.a11, d13040148}.\\

On the other hand, European wasps were much less often encountered against natural backgrounds than the pollinators. Manual classification of wasp backgrounds revealed only a $70\%$ encounter rate against natural microhabitats with a $63\%$ rate determined by automated classification. This confirms the wasps' long-recognised pest status in Australia \citep{doi:10.1080/03014223.1991.10422843}, 
with a diet that has come to include not only ``natural'' resources, but also human food and waste, especially carbohydrates \citep{goodall2001european}. 
This diet ensures humans encounter European wasps even in highly anthropogenic areas, and supports the reduction of access to human food and waste as a viable way to reduce encounters with these stinging and aggressive insects \citep{10.1371/journal.pone.0181397}. 
Future work would be valuable to unravel the extent to which wasps utilise natural and anthropogenic microhabitats differentially in natural, rural and urban regions.
\newline

The deep learning approach provided quick, reliable, automatic classification of image backgrounds. ResNet-50 classified with a very high test accuracy of $97.4\%$ (Section \ref{res:classifier_model}) using only the CPU of a basic laptop computer. Our trained ResNet-50 + SVM classifier took approx. $90 s$ to classify $500$ image backgrounds on this modest hardware, evidence of the relatively low computational requirements of the method.\\

It is interesting to consider the discrepancies between the results of our manual and automated methods for background classification (Fig. \ref{fig:misclassification}). As noted, ground and wood were the backgrounds where classifier improvement would be most fruitful. Some ground cover, such as grit and stones embedded in finer soil, can be recognised by humans as natural but further classifier training would improve the software's ability to distinguish this from anthropogenic conglomerate of concrete. Textured bark and naturally or roughly split wood backgrounds were sometimes classified as artificial by the software - their similarity to fence palings and weathered outdoor furniture is obvious.  But even in these cases the deviation of the automated method from the manual classification was very low, did not impact on the results of our study statistically, and therefore did not warrant further work for this project.\\

The difference in the discrepancy of background classification of ground and wood between manual and automated methods across the three target species (flies seemingly showed the lowest discrepancy, followed by bees and wasps), reflects the frequency with which these insects appeared against each background – an insect that isn't photographed on the ground for instance, would not raise a histogram bar in this category.\\

In Section \ref{manual-class}, we discussed how we discarded watermarked images from the ALA data before processing. If watermarked images provided important data, automated software could be applied to remove the water marks \citep{Liu_2021_WACV}. A further expansion of our method relates to non-insect studies. In the case of insects, most useful photographs are close-ups. But our method could potentially operate on larger animals as long as ``hybrid'' backgrounds were carefully handled; the larger the animal in the image, the more likely diverse elements will appear in the camera frame. It would then be essential to subdivide image backgrounds into regions to determine which element was most relevant to an animal's interaction with the location as specified in a particular research project. For instance, if an image showed a kangaroo eating grass (natural) whilst standing in front of an oncoming motor vehicle (anthropogenic), perhaps the food source is most relevant to the animal's use of the locality. But perhaps this would not be true if the study was unravelling why Australian native animals end up as road-kill. Along similar lines, in our study all images of insect hives were classified as natural. Some images may have shown artificial hives though - it can be impossible to tell from some images. In our dataset insect hives were few ($0.09\%$ of the total images), and the significant amount of effort to elucidate their location was deemed insufficiently valuable to explore in the current study. This may well be a fruitful avenue for future work exploring insect nest and hive locations specifically.\\

Microhabitats provide insects the resources to occupy an area that might, at a broader scale, seem inhospitable to them. The availability and explicit provision of suitable microhabitats for desirable insects within our built environments is increasingly important under a changing global environment and expanding urbanisation and industrial agriculture. And the reverse is also true. We would be wise to reduce access of pest insects to resources they depend on in our built environments to reduce the potential for humans to encounter dangerous or aggressive pests. Insect microhabitat can, and should, continue to be studied in traditional ways using on-the-ground field ecologists and manual data collection. Only in this way will a deep understanding of insect interactions with their environment be obtained. However, these painstaking, labour intensive, costly but essential projects can be supplemented by the image analysis procedures detailed here. In fact, traditional ecological projects could support our analysis approach by explicitly capturing relevant insect images for analysis.

\section{Conclusion}
The backgrounds of insect images provide important information from which to study the microhabitat use of individual insect species. We have developed new algorithmic procedures and an image analysis pipeline to extract insect pixels from an image, and successfully shown how manual approaches and automated machine learning can be used to classify microhabitat appearing in image backgrounds into broad natural and anthropogenic classes. Our findings suggest that a deep-learning based automated classifier model can quickly and accurately classify insect image backgrounds as benchmarked against manual classifications.
\newline

We found pollinating drone flies and honey bees had clear signatures in the image record of their use of natural microhabitats. This supports the findings of traditional studies that these insects require the provision of natural resources within built environments for their continued well-being and survival. This need set the drone flies and honey bees apart clearly from European wasps that are scavengers known to utilise human-provided resources in anthropogenic microhabitat. The wasps were, by contrast, often documented in the image data against anthropogenic backdrops. This finding supports the idea that depriving European wasps of access to human food, waste, and other anthropogenic features they depend on, would be one way to reduce encounters with these insects in built environments.

\section{Acknowledgement}
SSR was supported by the Faculty of Information Technology International Postgraduate Research Scholarship, Monash University, Australia.

\section{Author contributions}
All authors devised the idea. SSR implemented and conducted the methodology, and administered the analysis. SSR wrote the original draft of the manuscript. SSR and AD prepared the figures. All authors edited and contributed critically to the drafts and provided final approval for publication. AD and RT provided supervision and acquired the funding for this project.

\section{Declaration of competing interest}
The authors have no conflict of interest to declare.

\bibliographystyle{plainnat}
{\footnotesize
\bibliography{sample}}

\end{document}